\documentclass[12pt]{iopart}
\usepackage{graphicx}

\begin{document}

\title[]{The complex multiferroic phase diagram of Mn$_{1-x}$Co$_{x}$WO$_4$}

\author{K.-C. Liang$^1$, Y.-Q. Wang$^1$, Y. Y. Sun$^1$, B. Lorenz$^1$,  F. Ye$^2$, J. A. Fernandez-Baca$^{2,3}$, H. A. Mook$^2$, and C. W. Chu$^{1,4}$}

\address{$^1$ TCSUH and Department of Physics, University of Houston, Houston, TX 77204, USA}
\address{$^2$ Neutron Scattering Science Division, Oak Ridge National Laboratory, Oak Ridge, TN 37831-6393, USA}
\address{$^3$ Department of Physics and Astronomy, The University of Tennessee, Knoxville, TN 37996-1200, USA}
\address{$^4$ Lawrence Berkeley National Laboratory, 1 Cyclotron Road, Berkeley, CA 94720, USA}

\begin{abstract}
The complete magnetic and multiferroic phase diagram of Mn$_{1-x}$Co$_{x}$WO$_4$ single crystals is investigated by means of magnetic, heat capacity, and polarization experiments. We show that the ferroelectric polarization $\overrightarrow{P}$ in the multiferroic state abruptly changes its direction twice upon increasing Co content, x. At x$_{c1}$=0.075, $\overrightarrow{P}$ rotates from the $b-$axis into the $a-c$ plane and at x$_{c2}$=0.15 it flips back to the $b-$axis. The origin of the multiple polarization flops is identified as an effect of the Co anisotropy on the orientation and shape of the spin helix leading to thermodynamic instabilities caused by the decrease of the magnitude of the polarization in the corresponding phases. A qualitative description of the ferroelectric polarization is derived by taking into account the intrachain ($c-$axis) as well as the interchain ($a-$axis) exchange pathways connecting the magnetic ions. In a narrow Co concentration range (0.1$\leq$x$\leq$0.15), an intermediate phase, sandwiched between the collinear high-temperature and the helical low-temperature phases, is discovered. The new phase exhibits a collinear and commensurate spin modulation similar to the low-temperature magnetic structure of MnWO$_4$.
\end{abstract}

%\pacs{75.30.Kz, 75.50.Ee, 77.80.-e}

\maketitle

\section{Introduction}
The coexistence of different fundamental orders in solids and their mutual interaction have attracted attention of physicists and chemists for more than a century. Magnetoelectricity and the coupling between magnetism and ferroelectricity have inspired experimentalists and theorists because of many novel physica phenomena that arise from the complex cross correlations between magnetic and electric orders as well as the potential of these multiferroic materials for future applications in form of magnetoelectric sensors or new memory elements \cite{fiebig:05,spaldin:05,tokura:07}.

With the discovery of ferroelectricity induced as a secondary order parameter by an inversion symmetry breaking magnetic order in TbMnO$_3$ \cite{kimura:03}, the interest in multiferroic magnetoelectric compounds has been revived and the research in the field has made significant progress since then. The transverse helical spin structure is one of the candidates to induce ferroelectricity whenever the spin lattice coupling is strong enough to cause the cooperative ionic displacements resulting in a macroscopic electrical polarization. Some typical examples of spin-helical multiferroics are TbMnO$_3$ \cite{kenzelmann:05}, Ni$_3$V$_2$O$_8$ \cite{lawes:05}, and MnWO$_4$ \cite{arkenbout:06,taniguchi:06}. The important role of helical magnetic orders in inducing ferroelectricity in multiferroics has been theoretically studied within the Ginzburg-Landau model \cite{kenzelmann:05, mostovoy:06} or microscopic theories \cite{katsura:05,sergienko:06}.

The origin of helical magnetic orders is frequently found in magnetic frustration, either due to geometric constraints (for example antiferromagnetic interactions on a triangular lattice) or because of competing exchange interactions. In frustrated systems, there are several magnetically ordered phases that are very close in energy and they compete for the ground state. Small perturbations can therefore control the magnetic phases and stabilize one or the other of the competing orders. This results in an extreme sensitivity of most multiferroic materials with respect to magnetic or electric fields \cite{taniguchi:06,higashiyama:04,hur:04,seki:08,sagayama:08}, external pressure \cite{delacruz:07,chaudhury:08b,delacruz:08b,chaudhury:07}, and ionic substitutions \cite{seki:07,kanetsuki:07}. While the magnetic field couples to the ordered moments affecting their orientation and changing the magnetic structure, the application of external pressure compresses the lattice and tunes the microscopic exchange interactions and anisotropy parameters. Whereas the control of microscopic interactions via the application of pressure is indirect and limited by the lattice compressibility, the substitution of the magnetic ions by other ions with different spins, exchange couplings, and anisotropy constants allows for a more targeted control of the magnetic interactions.

The ideal multiferroic compound for investigating the effects of substitutions is MnWO$_4$. Also known as the mineral H\"{u}bnerite, MnWO$_4$ crystallizes in a monoclinic structure (space group P 2/c). At low temperature, several magnetic phase transitions and multiferroic properties have been reported \cite{arkenbout:06,taniguchi:06}. The spins of the only magnetic ion, Mn$^{2+}$, form different frustrated orders upon decreasing temperature because of competing magnetic exchange interactions. A sinusoidal magnetic order with an incommensurate modulation defined by the vector $\overrightarrow{Q}_{3}=(-0.214,1/2,0.457)$ sets in at T$_N$=13.5 K \cite{lautenschlager:93}. The Mn-spins of this AF3 phase are collinear and they are confined to the $a-c$ plane at an angle of 34$^{\circ}$ with the $a-$axis. At T$_C$=12.6 K, the spins form a non-collinear order by tilting partially toward the $b-$axis. The magnetic order in this AF2 phase is described by a spin-helix with the same propagation vector, $\overrightarrow{Q}_{2}$=$\overrightarrow{Q}_{3}$. Since the helical order breaks the spatial inversion symmetry, the AF2 phase is also ferroelectric with a polarization vector pointing along the $b$-axis, in perfect agreement with the spin-current model \cite{katsura:05} and the Ginzburg-Landau theory \cite{mostovoy:06}. Only at lower temperature, at T$_L$=7.5 K, the magnetic order locks into a commensurate structure described by the vector $\overrightarrow{Q}_{1}=(-0.25,0.5,0.5)$. The order in this AF1 phase shows the highly frustrated $\uparrow\uparrow\downarrow\downarrow$ spin sequence which preserves the spatial inversion symmetry. Therefore, the AF1 phase is paraelectric.

The complex phase sequence is a direct consequence of a high degree of frustration in the magnetic subsystem. The frustration originates from competing exchange interactions \cite{lautenschlager:93} and results in several magnetic states having almost the same energy and competing for the ground state. Recent neutron scattering experiments have revealed the long-range character of the magnetic interactions involving up to 11 different exchange coupling constants and the spin anisotropy to explain the low-energy magnetic excitations \cite{ye:11}. Replacing the magnetic Mn ion by other transition metal ions with different spin, exchange coupling, and anisotropy parameters will tune the various magnetic and multiferroic phases and, possibly, reveal new physical phenomena in frustrated systems. MnWO$_4$ is a good candidate since it forms stable compounds when Mn is completely replaced by Fe \cite{kleykamp:80}, Co, Ni \cite{weitzel:70}, or Zn \cite{takagi:81}. A uniform solid solution of MnWO$_4$ with (Fe,Co,Ni,Zn)WO$_4$ is expected to form over the whole concentration range because all ternary compounds are isostructural with relatively small variations of the lattice constants.

The substitution of Fe$^{2+}$ (spin 2) for Mn$^{2+}$ (spin 5/2) has shown an interesting magnetic phase diagram \cite{garciamatres:03} with coexisting phases at some critical Fe concentrations \cite{stuesser:01}. The multiferroic properties of Mn$_{1-x}$Fe$_x$WO$_4$ have been investigated earlier \cite{chaudhury:08} and it was shown that the substitution of more than 4 \% Fe suppresses the multiferroic AF2 phase. On the contrary, the replacement of nonmagnetic Zn$^{2+}$ in Mn$_{1-x}$Zn$_x$WO$_4$ did stabilize the AF2 phase as the ground state and the commensurate ($\uparrow\uparrow\downarrow\downarrow$) AF1 phase was completely suppressed \cite{meddar:09,chaudhury:11}. A similar effect was also observed in the Co substituted compound, Mn$_{1-x}$Co$_x$WO$_4$, based on magnetic and neutron scattering experiments of polycrystalline samples \cite{song:09}. The neutron scattering data further suggested the rotation of the FE polarization from the $b-$axis toward the $a-$axis at higher Co content. A sizable $a-$axis component of the polarization was indeed recently found in a single crystal of Mn$_{0.9}$Co$_{0.1}$WO$_4$ \cite{song:10,olabarria:12}. At slightly higher doping (15 \% Co), however, the only component of the FE polarization was aligned with the $b-$axis \cite{chaudhury:10}. These seemingly conflicting results indicate an extreme complexity of the multiferroic phase diagram of Mn$_{1-x}$Co$_x$WO$_4$ and warrant further exploration of single crystals of various Co concentrations.

In order to study the multiferroic and magnetic states and the polarization rotations between $b-$ and $a-$axes of Mn$_{1-x}$Co$_x$WO$_4$ in more detail, we have investigated the full doping range of Mn$_{1-x}$Co$_x$WO$_4$ between x=0 and x=0.3 through magnetic, dielectric (polarization), and heat capacity measurements. We show that, at zero magnetic field and upon increasing Co concentration, the ferroelectric polarization rotates twice, from $\overrightarrow{b}$ into the $a-c$ plane at x$\simeq$ 0.075 and back to $\overrightarrow{b}$ for x$>$ 0.135, limiting the multiferroic phase with the polarization perpendicular to $\overrightarrow{b}$ to a narrow region in the phase diagram.

\section{Experimental}
A large number of single crystals of Mn$_{1-x}$Co$_x$WO$_4$ with fourteen different compositions x between 0 and 0.3 have been grown in a floating zone optical furnace. Powder X-ray diffraction was used to check the phase purity of the polycrystalline feed rod before the crystal growth. No impurity phases could be detected within the resolution of the spectra. The chemical composition and the Co content of the single crystals were verified by energy-dispersive X-ray (EDX) measurements testing up to 15 different spots of a single crystal. For the 2 \% Co sample the actual Co content was 2.6 \%, slightly higher than the nominal substitution level. For all other samples the Co content was close to the nominal composition within the error limits of the EDX data. In the following sections, we will use the nominal composition to distinguish between different substitution levels.

The crystals cleave easily and expose the $a-c$ plane which was used as the contact area for $b-$axis polarization measurements. For measurements along other directions ($a-$ and $c-$axes), the crystals were cut accordingly. Single crystal Laue diffraction was employed to determine and confirm the crystalline orientations.

For dielectric and ferroelectric polarization measurements, small plate-like samples were cleaved or cut from the bigger single crystal. Silver paint was used for electrical contacts. The contact area was about 10 to 15 mm$^2$ and the sample thickness was typically less than 0.5 mm. The sample was mounted on a home-made probe inserted into the Physical Property Measurement System (Quantum Design) for precise control of temperature and the speed of temperature change. The spontaneous polarization was measured through the pyroelectric current method using the K6517A electrometer (Keithley). The experimental protocol was chosen to diminish any artificial contributions to the current as follows: The sample was cooled from 30 K (above T$_N$) in an electric field of about 3 kV/cm to low temperature while measuring the current. After releasing the applied voltage and shortening the contacts for several minutes at the lowest temperature, the pyroelectric current was measured upon increasing temperature. The rate of temperature change was 1 K/min in all experiments. The polarization is than calculated by integrating the current from high temperature (in the paraelectric state) to low temperatures. The cooling data for the polarization have been found consistent with the (zero electric field) warming data, but they reveal large thermal hysteresis effects in some part of the phase diagram.

The bias electric field during cooling (3 kV/cm) has to be high enough to ensure the alignment of the FE domains. This was verified by measuring the FE polarization after cooling in a reduced electric field of 2 kV/cm resulting in the same magnitude of the FE polarization. Some experiments had been conducted in cooling in a negative (sign-reversed) electric field and the obtained polarization was of equal magnitude but of opposite sign, as expected in a FE state.

Magnetization measurements have been conducted in a commercial magnetometer (Magnetic Property Measurement System, Quantum Design). The low-field data were typically taken in a magnetic field of 200 Oe. The heat capacity was measured using the relaxation method implemented in the heat capacity option of the PPMS.

\section{Results and Discussion}
\subsection{Magnetic susceptibility of Mn$_{1-x}$Co$_x$WO$_4$}
The small amount of Co doping of only 0.02 suppresses the low-temperature paraelectric AF1 phase. This is revealed in magnetization data shown in Fig. 1a. The transitions into the sinusoidal (T$_N$) and the helical magnetic phases (T$_C$) are resolved in anomalies of the $b-$axis magnetic susceptibility, as indicated in Fig. 1a. The phase transition from the helical AF2 to the commensurate AF1 phase (T$_L$) is only detected in MnWO$_4$ as a sharp increase of $\chi_b$(T) at T$_L$. In Mn$_{0.98}$Co$_{0.02}$WO$_4$, however, this transition is already missing. The quick suppression of the AF1 phase extends the ferroelectric AF2 phase to the lowest temperature. This result is consistent with similar conclusions derived from the study of polycrystalline Mn$_{1-x}$Co$_x$WO$_4$ \cite{song:09}. An analogous doping effect on the stability of the AF1 and AF2 phases has been observed recently in single crystals of Mn$_{1-x}$Zn$_x$WO$_4$ \cite{chaudhury:11}. For 0.02$\leq$x$\leq 0.05$, $\chi_b$(T) changes very little (different curves in Fig. 1a are vertically offset) but T$_N$ and T$_C$ are slightly decreasing.

With increasing x between 0.075 and 0.15, however, $\chi_b$(T) changes significantly below T$_C$. The $b-$axis magnetic susceptibilities in this range of Co substitutions are shown in Fig. 1b. The relatively sharp decrease of $\chi_b$ below T$_C$ disappears above x=0.075 and it changes into a plateau-like feature that dominates the low-temperature behavior of the susceptibility. The low-T plateau exhibits additional anomalies in form of small step-like changes of $\chi_b$(T) for x$\geq$0.12 suggesting additional transitions or changes of the magnetic orders and a more complex magnetic phase diagram in this doping range (details are discussed below).

With further increasing x$\geq$0.17, however, the T-dependence of $\chi_b$ again changes qualitatively, as shown in Fig. 1c. While the Ne\'{e}l transition is clearly defined by a sharp slope change of $\chi_b$(T), the susceptibility continues to increase to lower T until it suddenly decreases sharply below a second transition temperature, T$_C$. The T-dependence of $\chi_b$ below T$_C$ is similar to the susceptibility data shown in Fig. 1a for the low-doping range. The two anomalies labeled as T$_N$ and $T_C$ separate two magnetic phases. With the results from powder neutron scattering \cite{song:09} and our earlier results for x=0.15 \cite{chaudhury:10}, the phase between T$_N$ and T$_C$ can be identified as the AF4 phase with the commensurate magnetic modulation corresponding to $\overrightarrow{Q}_{4}=(0.5,0,0)$, similar to the magnetic phase observed in Fe-substituted MnWO$_4$ above 12.5 \% iron content \cite{garciamatres:03}.

\begin{figure}
\begin{center}
\includegraphics[angle=0,width=5in]{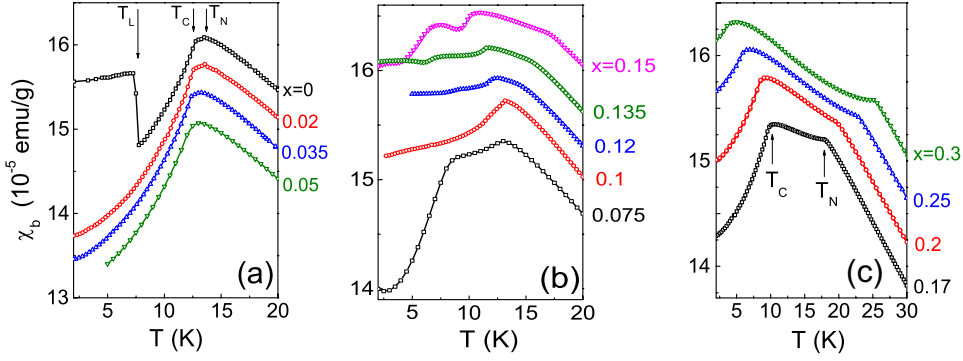}
\caption{$b-$axis magnetic susceptibility of Mn$_{1-x}$Co$_x$WO$_4$. The arrows indicate the critical temperatures T$_N$ (AF3, AF4), T$_C$ (AF2), and T$_L$ (AF1). Different curves are vertically offset for better clarity by 0.35, 0.3, and 0.4 units in (a), (b), and (c), respectively.}
\end{center}
\end{figure}

\subsection{Heat capacity of Mn$_{1-x}$Co$_x$WO$_4$}
The various phase transitions between different magnetic phases are reflected in distinct peaks or anomalies of the heat capacity, C$_p$(T). For example, the heat capacity of the undoped MnWO$_4$ exhibits pronounced peaks at all three magnetic transitions from the paramagnetic phase to AF3, from AF3 to AF2, and from AF2 to AF1 \cite{arkenbout:06}. Our single crystal data for x=0 are included in Fig. 2a, top curve. For 0.02$\leq$x$\leq0.1$, the heat capacity data, shown in Fig. 2a, reveal only two peaks between 11 and 14 K, representing the transitions from the paramagnetic to the sinusoidal and subsequently to the helical phases, respectively, similar to MnWO$_4$ (note that the AF1 phase is already suppressed in this doping range).

Upon increasing x, the onset of magnetic order at T$_N$ is accompanied by a step-like increase of C$_p$ and a large peak of C$_p$ develops near 11 K together with a small third peak at lower temperature between x=0.12 and x=0.15 (Fig. 2b). For x$\geq$0.17, the low-temperature peak of C$_p$(T) disappears and only two anomalies remain, indicating that two magnetic phases dominate the phase diagram in this doping range (see Fig. 2b). The two transition temperatures exhibit a large x-dependence with the higher transition (T$_N$) quickly increasing and the second transition (T$_C$) shifting to lower temperatures.

The x-dependencies of the low-temperature properties of $\chi_b$ and C$_p$ suggest that there are at least three doping ranges (denoted by I, II, and III) in the phase diagram of Mn$_{1-x}$Co$_x$WO$_4$ between x=0 and x=0.3 that are distinguished by their characteristic magnetic properties and thermodynamic anomalies, as shown in Figs. 1 and 2, respectively. In ranges I (0.02$\leq$x$\leq$0.05) and III (0.17$\leq$x$\leq$0.3) only two transitions are observed, separating two phases with different magnetic structures. In range II, the phase sequence upon decreasing temperature appears to be more complex involving more than two magnetic orders. Therefore the question arises which of the various phases is ferroelectric (or multiferroic). This can only be decided by measurements of the ferroelectric polarization.

\begin{figure}
\begin{center}
\includegraphics[angle=0,width=4in]{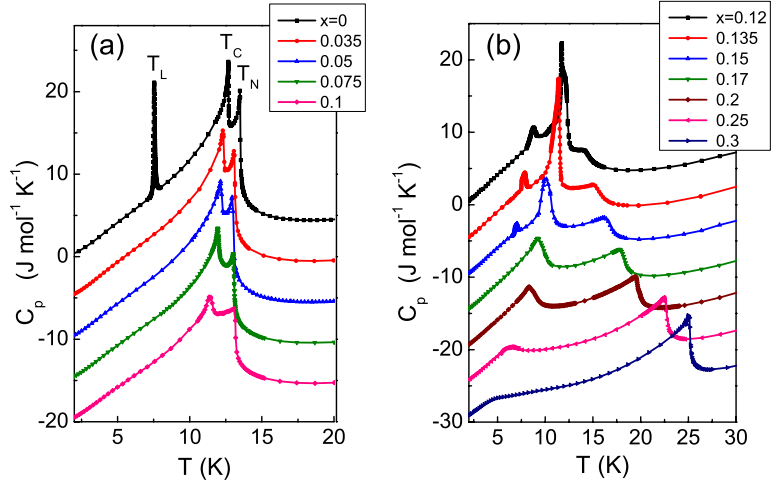}
\caption{Molar heat capacity of Mn$_{1-x}$Co$_x$WO$_4$. Different curves are vertically offset for better clarity by -0.5 units with reference to the x=0 and x=0.12 data in (a) and (b), respectively. The labels in (a) for x=0 indicate the peaks associated with the magnetic transitions at T$_N$ (AF3), T$_C$ (AF2), and T$_L$ (AF1).}
\end{center}
\end{figure}

\subsection{Ferroelectricity in Mn$_{1-x}$Co$_x$WO$_4$}
The ferroelectric polarization $\overrightarrow{P}$(T) in the AF2 phase of MnWO$_4$ is directed along the monoclinic $b-$axis \cite{arkenbout:06}. This is in accordance with the symmetry and the prediction of the spin current model, with the $b-$axis lying in the plane of the spin helix and assuming that the major contribution arises from the average nearest neighbor spin-spin exchange oriented along the $c-$axis. For x=0.1, however, it was shown that the polarization had rotated by 90$^\circ$ into the $a-c$ plane \cite{song:10}. A smaller $c-$axis component found in Mn$_{0.9}$Co$_{0.1}$WO$_4$ was attributed to the magnetic interaction along the $a-$axis between spins of different chains \cite{song:10,olabarria:12}.
In order to account for all possible orientations of $\overrightarrow{P}$(T) for 0$\leq$x$\leq$0.3, we have measured for each composition the polarization components along $a$, $b$, and $c$.

For Co substitution levels up to 0.05, $\overrightarrow{P}$(T) was found to be aligned with the $b-$axis (Fig. 3a). The undoped MnWO$_4$ shows the sharp drop of $\overrightarrow{P}$(T) to zero at the lock-in transition into the AF1 phase, in agreement with earlier reports \cite{arkenbout:06,taniguchi:06}. With x increasing above 0.02, however, the paraelectric AF1 phase is quenched and the ferroelectric AF2 phase extends to the ground state. The low-temperature (measured at 5 K) polarization decreases with x and reaches about 50 \% of the value of MnWO$_4$ at x=0.05. The reference value for MnWO$_4$ at 5 K, P$_b\simeq$60 $\mu$C/m$^2$, was derived from data measured at a magnetic field of 3 T which also suppresses the paramagnetic AF1 phase \cite{arkenbout:06}. The ferroelectric critical temperature T$_C$ decreased slightly from 12.8 K (x=0) to 12.3 K (x=0.05). At the upper limit of this Co concentration interval, for x=0.042 and 0.05, a small component of the polarization along the $a-$axis was detected. The size of P$_a$ was more than one order of magnitude smaller than P$_b$. This small P$_a$ could be an effect of a minute misalignment of the crystal or it could indicate the onset of the polarization flop from $b$ into the $a-c$ plane. Our measurements of the ferroelectric polarization of Mn$_{0.95}$Co$_{0.05}$WO$_4$ in magnetic fields \cite{liang:12} are consistent with the onset of a smooth rotation of the polarization (and the spin helix) at doping levels as low as 0.042.

\begin{figure}
\begin{center}
\includegraphics[angle=0,width=6in]{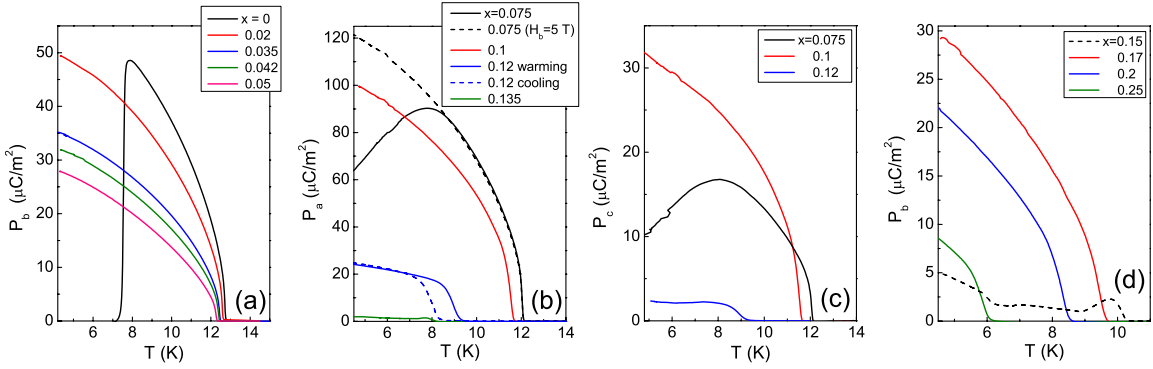}
\caption{Ferroelectric polarization of Mn$_{1-x}$Co$_x$WO$_4$. (a) Low doping range (x$\leq$0.05, P$_b$), (b, c) intermediate doping (0.075$\leq$x$\leq$0.135, P$_a$ and P$_c$), and (d) high doping range (x$\geq$0.15, P$_b$). For x=0.12, warming and cooling data are presented to show the thermal hysteresis. The dashed line for x=0.075 shows the maximum polarization achieved in magnetic field.}
\end{center}
\end{figure}

At slightly higher Co doping, from x=0.075 to x=0.135, the ferroelectric polarization does not show any significant $b-$axis component within the resolution of the measurement. Small values of P$_b$ detected in some measurements can be due to a minute misalignment (typically less than 3$^\circ$) from the perfect orientation. However, a large polarization was found along the $a-$axis accompanied by a smaller component along $c$, as shown in Figs. 3b and 3c, respectively. The magnitude of P$_c$ is too large to be explained by a misalignment and it has to be understood as an intrinsic effect. Since the nearest neighbor intrachain magnetic couplings are oriented along the $c-$axis, they cannot generate a polarization along $c$ and the interchain magnetic interactions along $a$ have to be considered. This is discussed in more detail in Section 3.5.

The magnitudes of P$_a$ and P$_c$ at 5 K decrease quickly with increasing x and nearly vanish at x=0.135. For x=0.075, both P$_a$ and P$_c$ pass through a maximum at 9 K and decrease upon further cooling. This unusual behavior can be attributed to a temperature dependent reorientation of the helical plane from the optimal position in the $a-c$ plane resulting in a decrease of the polarization below 9 K. A magnetic field oriented along the $b-$axis was shown to stabilize the $a-$axis component of $\overrightarrow{P}$(T) (as for example in Mn$_{0.95}$Co$_{0.05}$WO$_4$ \cite{liang:12}). Indeed, P$_a$ of Mn$_{0.925}$Co$_{0.075}$WO$_4$ measured in a field H$_b$=5 Tesla rises continuously and reaches the maximum value of more than 120 $\mu$C/m$^2$ at 5 K, as shown by the dashed line in Fig. 3b. It should be noted that this value is the highest one measured in any doped or undoped MnWO$_4$ compound and it exceeds the maximum $b-$axis polarization of MnWO$_4$ by a factor of two. It is also interesting that this maximum of $\overrightarrow{P}$ appears right at the critical x at which the rotation of the polarization from the $b-$axis into the $a-c$ plane happens.

The data for x=0.1 are in quantitative agreement with recent reports \cite{song:10,olabarria:12}. Upon further increasing x, $\overrightarrow{P}$(T) decreases quickly and the critical temperature of the ferroelectric phase shows a sharp decrease from 11.3 K (x=0.1) to 7.6 K (x=0.135). A significant thermal hysteresis of more than 1 K develops above x=0.1, as demonstrated by the warming and cooling data for x=0.12, included in Fig. 3b. The values of T$_C$ of the ferroelectric phase match perfectly with the temperatures of the small peak of the heat capacity (Fig. 2b). Near the upper limit of region II of the phase diagram (x=0.135), the polarization at 5 K dropped already to below 2 $\mu$C/m$^2$. The fast decrease of $\overrightarrow{P}$ with increasing x and the fact that no sizable component of $\overrightarrow{P}$ was observed along $\overrightarrow{b}$ indicate a major change of the spin helix in a narrow doping range.

Increasing the Co content x above 0.135 results in another flop of the ferroelectric polarization from the $a-c$ plane back to the $b-$axis with a sudden increase of its magnitude. Data for x$\geq$0.15 are shown in Fig. 3d. The results for x=0.15, previously reported by us \cite{chaudhury:10}, are included as the dashed line. The recovery of the $b$-axis polarization up to 30 $\mu$C/m$^2$ at 5 K suggests another sudden rotation of the spin helix in Mn$_{1-x}$Co$_x$WO$_4$ at a critical doping of 0.15. No sizable $a-$ or $c-$axis components of $\overrightarrow{P}$ could be detected in this higher doping range.

\begin{figure}
\begin{center}
\includegraphics[angle=0,width=3in]{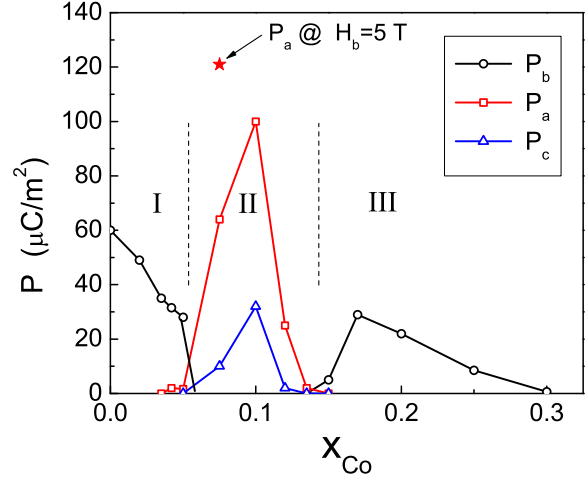}
\caption{Low-temperature (5 K) values of the ferroelectric polarization of Mn$_{1-x}$Co$_x$WO$_4$ showing the separation into three ferroelectric phases I, II, and III, with different orientations of $\overrightarrow{P}$. For x=0, the value of P was extrapolated to low temperature based on data in magnetic fields suppressing the AF1 phase. For x=0.135 the maximum value obtained at 5 Tesla is shown as a star.}
\end{center}
\end{figure}

The x-dependency of the ferroelectric polarization is summarized in Fig. 4 showing the values of $\overrightarrow{P}$ at 5 K vs. Co doping x. The two instabilities near x$_{c1}$=0.075 and x$_{c2}$=0.135 are clearly represented and it is obvious that the rotation of $\overrightarrow{P}$ happens when its magnitude drops below a critical value. Above x=0.3, $\overrightarrow{P}$ decreases quickly to zero and the helical magnetic phase is quenched completely. The collinear AF4 phase with the characteristic spin modulation defined by $\overrightarrow{Q}_{4}=(0.5,0,0)$ becomes the ground state. This is consistent with the AF4 magnetic structure that is realized for x=1, i.e. in CoWO$_4$ \cite{forsyth:94}.

\subsection{Magnetic and multiferroic phase diagram of Mn$_{1-x}$Co$_x$WO$_4$}
The polarization measurements discussed above define the ferroelectric (or multiferroic) phases of Mn$_{1-x}$Co$_x$WO$_4$. However, other magnetic phases, e.g. the paraelectric AF1, AF3, or AF4 phases, have to be defined through their characteristic properties and the magnetic as well as thermodynamic anomalies at the phase boundaries (see sections 3.1 and 3.2 above). Combining all experimental data, we can completely resolve the magnetic and multiferroic phase diagram as shown in Fig. 5. The phase assignment of different ordered structures used in Fig. 5 is based on previous \cite{song:09} and most recent \cite{ye:12} neutron scattering results.

\begin{figure}
\begin{center}
\includegraphics[angle=0,width=3in]{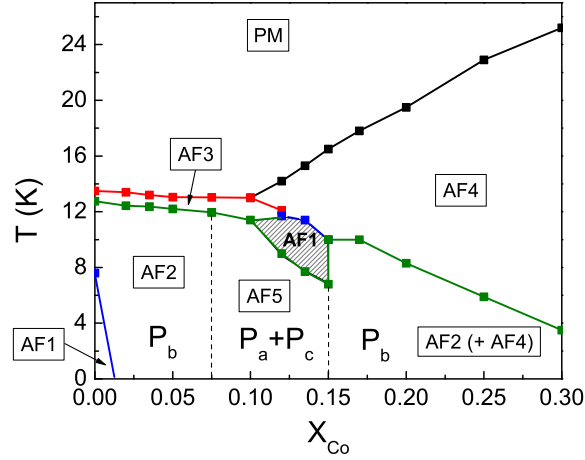}
\caption{Magnetic and multiferroic phase diagram of Mn$_{1-x}$Co$_x$WO$_4$. Phase boundaries are derived from magnetic, heat capacity, and polarization measurements. The preliminary phase assignments are based on recent neutron scattering reports \cite{ye:12}. The shaded area indicates the intermediate paraelectric AF1 phase, as discussed in the text.}
\end{center}
\end{figure}

Some features of the phase diagram, such as the quick suppression of the AF1 phase and the overall onset of magnetic order at T$_N$, are in agreement with the results from powder neutron scattering experiments \cite{song:09}. The Ne\'{e}l temperature decreases slightly with x for x$\leq$0.1, the first magnetic structure in this doping range is the sinusoidal AF3 phase. However, T$_N$ increases significantly with x$>$0.1 when the commensurate AF4 phase replaces the incommensurate AF3 phase. The transition from the sinusoidal AF3 to the helical AF2 phase with the polarization along the $b-$axis extends to x$\simeq$0.075. Between x$_{c1}$=0.075 and x$_{c2}$=0.135, however, the ferroelectric polarization is constrained to the $a-c$ plane exclusively (AF5 phase).

It should also be noted that for 0.12$\leq$x$\leq$0.15 three magnetic phase transitions are observed in magnetic data (Fig. 1) and heat capacity (Fig. 2). A third phase (shaded area in Fig. 5), not reported so far, is sandwiched between the AF4 and AF5 phases. According to the polarization measurements, this phase is paraelectric. The transition into this phase is accompanied by the large peak of the heat capacity, as shown in Fig. 2b. Recent detailed neutron scattering experiments \cite{ye:12} have revealed the magnetic structure of this phase, identifying it as a commensurate collinear magnetic structure with the wave vector of the low-temperature AF1 phase of MnWO$_4$. This is also consistent with our previous neutron data for x=0.15 \cite{chaudhury:10}. The details, however, have yet to be explored. The presence of this intermediate phase also explains the sudden decrease of the ferroelectric transition temperature in this range and its recovery above x=0.15.

With further increasing x ($>$0.15), only two phase transitions can be observed. At T$_N$, a commensurate magnetic order (AF4 phase) develops first, as reflected in the kink of $\chi_b$ (Fig. 1c) and the high-temperature peak of C$_p$ (Fig. 2b). The second transition at T$_C$ recovers the $b-$axis polarization, the magnetic structure therefore has to be noncollinear allowing for a polarization along $\overrightarrow{b}$ exclusively, similar to the AF2 phase at lower doping levels. P$_b$ as well as the critical temperature T$_C$ decrease quickly with x to P$_b$(2 K)=0.6 $\mu$C/m$^2$ and T$_C$=3.6 K, respectively, at x=0.3. The strong reduction of the polarization may be due to a decrease of the magnetic moments contributing to the spin helix, a possible change of the ellipticity, or to the increasing influence of the AF4 magnetic structure. A possible coexistence of both phases was suggested based on powder neutron scattering experiments \cite{song:09}.

\subsection{Discussion and possible origin of the multiple polarization flops in Mn$_{1-x}$Co$_x$WO$_4$}
The multiferroic state in MnWO$_4$ is the classical example of a polar structure induced by an inversion symmetry breaking helical magnetic order \cite{arkenbout:06,mostovoy:06,sagayama:08}. The strong coupling of the lattice to the spin helix, as revealed e.g. through significant anomalies of the thermal expansivity \cite{chaudhury:08c}, results in the polar distortion of the lattice and the ferroelectric state. Both, the microscopic theory as well as symmetry arguments relate the magnitude and the direction of the electrical polarization to the vector product of the term ($\overrightarrow{m}_i\times\overrightarrow{m}_j$) with the position vector $\overrightarrow{R}_{ij}$ connecting both magnetic moments at sites i and j along the propagation direction of the spin helix. Taking into account the specific structure of MnWO$_4$ with two spins per unit cell, the magnetic moment in the helical state can be expressed as: \cite{lautenschlager:93}

\begin{equation}
\overrightarrow{m}(\overrightarrow{R}_{i\alpha}) = \overrightarrow{m}_{\parallel}cos(\overrightarrow{q}\overrightarrow{R}_i+\Phi_\alpha)+\overrightarrow{m}_{\perp}sin(\overrightarrow{q}\overrightarrow{R}_i+\Phi_\alpha)
\end{equation}

$\overrightarrow{R}_{i\alpha}$ is the position vector to the Mn site $\alpha$ (=1, 2) in the unit cell i, $\overrightarrow{q}$ is the propagation vector of the spin helix defined as $q_i=2\pi/\lambda_i$ ($\lambda_i$ is the wavelenght of the spin modulation along the $i^{th}$ direction), and $\Phi_2$=$\Phi_1$+$q_zc/2$+$\pi$. The two perpendicular vectors $\overrightarrow{m}_{\parallel}$ and $\overrightarrow{m}_{\perp}$ are chosen to define the helical magnetization plane with m$_\parallel$ and m$_\perp$ being equal to the long and short half axis of the magnetic ellipse, respectively.

To derive a general expression for the ferroelectric polarization we have to calculate the vector product of the two magnetization vectors at the pair of atomic sites coupled by the magnetic exchange interaction. The result for nearest neighbor spins within a chain along the $c-$axis is:

\begin{equation}
\overrightarrow{m}(\overrightarrow{R}_{i2})\times\overrightarrow{m}(\overrightarrow{R}_{i1})=m_{\parallel}m_{\perp}sin(q_zc/2)\overrightarrow{n}
\end{equation}

$\overrightarrow{n}$ is the normal vector of the helical plane (length 1), as indicated in Fig. 6.

As noted above, the coupling between spins exclusively within the chains ($c-$axis) cannot account for the observed $c-$axis component of the polarization in the intermediate doping range. Magnetic exchange between different chains has to contribute to the ferroelectric polarization. This is also suggested by the results of recent inelastic neutron scattering data \cite{ye:11} which lead to an estimate of the Heisenberg exchange constants for different pairs of spins along as well as in between the chains. The magnetic exchange between spins along the $b-$direction was generally very weak; however, sizable exchange constants had been determined for neighboring spins along the $c-$ as well as $a-$axes. Therefore, it appears conceivable to include contributions to the polarization that arise from the spin-spin interactions along the $a-$axis:

\begin{equation}
\overrightarrow{m}(\overrightarrow{R}_{i+1,\alpha})\times\overrightarrow{m}(\overrightarrow{R}_{i\alpha})=m_{\parallel}m_{\perp}sin(q_xa)\overrightarrow{n}
\end{equation}

Here $\overrightarrow{R}_{i+1,\alpha}$ and $\overrightarrow{R}_{i\alpha}$ refer to two magnetic ions in neighboring chains displaced along the $a-$axis.

The contributions to the electrical polarization can now be expressed as the vector product of the expressions (2) and (3) with the corresponding average position vectors connecting the spins in each case. Using spherical coordinates, we express the normal vector $\overrightarrow{n}=sin(\Theta)cos(\varphi)\overrightarrow{e}_x+sin(\Theta)sin(\varphi)\overrightarrow{e}_y+cos(\Theta)\overrightarrow{e}_z$. $\Theta$ is the angle of $\overrightarrow{n}$ with the z-axis and $\varphi$ denotes the angle of the x-y plane projection of $\overrightarrow{n}$ with the x-axis. $\overrightarrow{e}_i$ are the unit vectors along the i-th cartesian coordinate (i=x,y,z). For simplicity, we will consider $\overrightarrow{e}_x$ parallel to $\overrightarrow{a}$ and $\overrightarrow{e}_z$ parallel to $\overrightarrow{c}$ (the angle between $\overrightarrow{a}$ and $\overrightarrow{c}$ is nearly 90$^\circ$), with $\overrightarrow{e}_y$ parallel to $\overrightarrow{b}$. The polarization in terms of $\Theta$ and $\varphi$ due to the intrachain exchange is:

\begin{equation}
\overrightarrow{P}^{(1)}=C^{(1)}m_{\parallel}m_{\perp}sin(q_zc/2)[-sin\Theta sin\varphi \overrightarrow{e}_x + sin\Theta cos\varphi \overrightarrow{e}_y]
\end{equation}

A similar expression is obtained for the interchain exchange along the a-axis:

\begin{equation}
\overrightarrow{P}^{(2)}=C^{(2)}m_{\parallel}m_{\perp}sin(q_xa)[-cos\Theta \overrightarrow{e}_y + sin\Theta sin\varphi \overrightarrow{e}_z]
\end{equation}

$C^{(1)}$ and $C^{(2)}$ are constants not depending on $\Theta$ or $\varphi$ with supposedly $|C^{(2)}|<|C^{(1)}|$, because of the larger distance between the chains and the weaker magnetic exchange. The signs of $q_z$ and $q_x$ determine the helicity of the spiral spin modulation.

In MnWO$_4$ the helical plane is perpendicular to the $a-c$ plane forming an angle of about 34$^\circ$ with the $a-$axis, corresponding to the spin easy axis in the collinear AF1 and AF3 phases \cite{lautenschlager:93}. Accordingly, the orientation of the helix is defined by $\Theta$=34$^\circ$ and $\varphi$=180$^\circ$. This results in a $b-$axis polarization with contributions from equation (4) as well as equation (5). Whether the two contributions to P$_b$ are additive or subtractive depends on the helicities of the spin spiral and the signs of the prefactors in equations (4) and (5), i.e. the microscopic details of the magnetic interactions.

\begin{figure}
\begin{center}
\includegraphics[angle=0,width=3in]{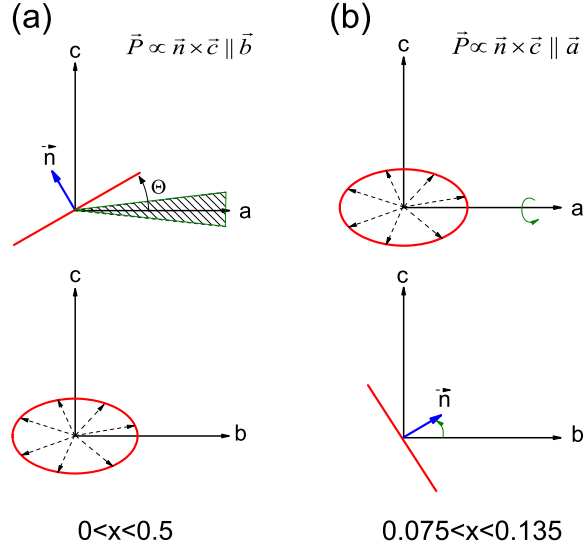}
\caption{Schematic diagram of the possible orientations (defined by the normal vector $\overrightarrow{n}$) of the spin helix in Mn$_{1-x}$Co$_x$WO$_4$ for x$\leq$0.05 (a) and 0.075$\leq$x$\leq$0.135 (b). The shaded area in (a) symbolizes the range where the helical plane comes close to the $a-$axis and the associated $b-$axis polarization is too small to stabilize the AF2 phase.}
\end{center}
\end{figure}

The dependence of the magnitude of the ferroelectric polarization, P$_b$, on the angle $\Theta$ was discussed by Sagayama et al. \cite{sagayama:08} taking into account only the contribution due to nearest neighbors along the $c-$axis, i.e. equation (4). However, because of the second contribution from equation (5), the angular dependence of P$_b(\Theta,\varphi)$ will be more complex.

In the following we will qualitatively discuss the complex x-dependent ferroelectric polarization as a consequence of the possible changes of the spin helix induced by the cobalt substitution. Using formulas (4) and (5), our focus will be on the angular dependence, i.e. the expressions in the square brackets, although the possible changes of the prefactors with x may cause additional changes of the magnitude of $\overrightarrow{P}$. The reorientation of the helical plane was recently observed for Co concentrations between 0.05 and 0.1 \cite{song:10,olabarria:12}. The driving mechanism, however, has yet to be revealed.

It appears conceivable that the change of the average spin anisotropy upon doping with cobalt plays a major role. The easy axis of the transition metal (T) spin in TWO$_4$ strongly depends on the T-ion \cite{weitzel:77}. For CoWO$_4$ it was determined that the angle of the easy axis with the $a-$axis is $\Theta=-46^\circ$ \cite{forsyth:94}, a change of 80$^\circ$ with respect to MnWO$_4$. It can therefore be expected that the Co substitution tends to rotate the easy axis and the helical plane in the multiferroic phase of Mn$_{1-x}$Co$_x$WO$_4$ into the direction of the spin easy axis of CoWO$_4$, as schematically indicated by the curved arrow in Fig. 6. This rotation was indeed observed in neutron scattering experiments for small x$\leq$0.05 \cite{song:09,song:10,ye:12}. The decrease of $\Theta$ from 34$^\circ$ (x=0) to 15$^\circ$ (x=0.05) results in a reduction of the ferroelectric polarization without changing its orientation, according to the data shown in Fig. 3a and Fig. 4.

The decrease of P$_b$, however, is energetically unfavorable since the multiferroic state gains energy from the coupling of the polarization to the magnetic order parameter. The energy gain originates from a third order term in the free energy expansion that is proportional to the polarization. This term, which has to be invariant with respect to the spatial inversion and time reversal operations, has the mathematical form of a Lifshitz invariant: $\gamma \overrightarrow{P}\cdot[\overrightarrow{M}(\nabla\cdot \overrightarrow{M})-(\overrightarrow{M}\cdot\nabla)\overrightarrow{M}]$ \cite{mostovoy:06}. The gain of free energy in the multiferroic state scales accordingly with the magnitude of the polarization. If the spin helix rotates closer to the $a-$axis ($\Theta<$15$^\circ$), the polarization as well as the energy gain in the multiferroic state will further decrease as long as the angle $\varphi$ remains unchanged. For energy reasons, we can therefore expect an instability at a concentration x$_{c1}$ when $\Theta$ decreases below a critical value, $\Theta_c$, and the multiferroic AF2 state cannot gain enough energy to be stable with respect to other competing states. The shaded area in Fig. 6a shows schematically the region of instability for the helical plane in this phase.

The multiferroic state can gain more energy through a flop of the helical plane allowing for larger values of the ferroelectric polarization. The flop can be visualized as a rotation of the helical plane toward the $a-c$ plane (for $\varphi$=90$^\circ$ and $\Theta$=90$^\circ$ the helical plane coincides with the $a-c$ plane). According to equations (4) and (5), this results in a rotation of $\overrightarrow{P}$ into the $a-c$ plane, where the two components P$_a$ and P$_c$ arise from equations (4) and (5), respectively. It is interesting to note that the angular dependence of both components is exactly the same implying a similar temperature or magnetic field dependence with different magnitudes of the polarization values (according to the different prefactors in equations (4) and (5)).

The $a-c$ polarized state is realized in the AF5 phase between x$_{c1}$=0.075 and x$_{c2}$=0.135 (Figs. 3b and 3c). The first signature of the polarization flop in the phase diagram of Mn$_{1-x}$Co$_x$WO$_4$ appears at x=0.075 (Fig. 3b). P$_a$ assumes the largest values above 8 K, exceeding even the values for x=0.1 (Fig. 3b and previous reports \cite{song:10,olabarria:12}). With further cooling, however, P$_a$ passes through a maximum and decreases again. This unusual behavior may indicate a temperature dependent tilt of the helical plane below 8 K away from the optimal orientation. The $c-$axis component exactly mimics the temperature dependence of P$_a$ at a smaller scale, as is expected from equations (4) and (5). Due to this low-temperature tilt, the maximum polarization cannot be obtained for x=0.075 in zero magnetic field. A magnetic field applied along the $b-$axis, however, reveals the maximum value of P$_a>$120$\mu C/m^2$ at 4.5 K, which appears to be the upper limit for the MnWO$_4$ system. The high-field data are included in Fig. 3b (dashed line) and Fig. 4 (star).

P$_a$ and P$_c$ decrease quickly between x$_{c1}$ and x$_{c2}$ and no sizable polarization component along the $b-$axis was observed. The data for x=0.1 are in good agreement with previous reports \cite{song:10,olabarria:12}. The observed decrease of the polarization may indicate another rotation of the spin helix, so that $\Theta<$90$^\circ$ and/or $\varphi<$90$^\circ$ and $sin\Theta sin\varphi<1$, as sketched schematically in Fig. 6b . However, the decrease of P$_a$ and P$_c$ to almost zero (P$_a$=2$\mu$C/m$^2$ for x=0.135 and T=5 K) without any sizable $b-$axis polarization is difficult to understand. If $\Theta$, $\varphi$ systematically decrease with x, both equations (4) and (5) are expected to also give rise to a significant component P$_b$, which was not experimentally observed. The question of whether the two contributions to P$_b$ accidentally may cancel each other has yet to be explored. Another possibility for the decrease of $\overrightarrow{P}$(x) in this section of the phase diagram could be an increase of the ellipticity ($m_\parallel \gg m_\perp$) of the spin helix with x, reducing the prefactor in equations (4) and (5). If $m_\perp$=0, the magnetic order is collinear and no polarization is allowed. The true nature of the magnetic order in the AF5 phase can only be revealed by detailed neutron scattering experiments.

Nevertheless, the decrease of P$_a$ and P$_c$ again diminish the stabilizing energy of the multiferroic state arising from the magnetoelectric coupling term in the free energy expansion. Therefore, another instability causes the polarization to flop back toward the $b-$axis (for x$\geq$0.15). A sizable value of the ferroelectric polarization along the $b-$axis is recovered and the multiferroic state is stabilized with regard to the nearly paraelectric state close to x=0.135. The further decrease of P$_b$ is driven by the competition with the commensurate and paraelectric AF4 magnetic structure that extends to lower temperatures and coexists with the helical phase below T$_C$. The phase boundary is right at x=0.15, as previous neutron scattering data of Mn$_{0.85}$Co$_{0.15}$WO$_4$ \cite{chaudhury:10} have revealed the coexistence of a sizable amount of AF5 phase with the arising AF2 phase (with $b-$axis polarization). From the above discussion it is now obvious that the AF5 phase at x=0.15 (which was shown to be the major phase at low temperatures) does not generate any significant polarization, but it reduces the AF2 phase to a minor phase. This explains the small value of P$_b$ (shown in Fig. 3c) as compared to the values for x$\geq$0.17.

\section{Summary}
We have investigated the complete multiferroic and magnetic phase diagram of Mn$_{1-x}$Co$_x$WO$_4$ for 0$\leq$x$\leq$0.3 through magnetization, heat capacity, and polarization measurements. With increasing x, the ferroelectric polarization abruptly changes direction at the critical Co concentrations of x$_{c1}$=0.075 ($\overrightarrow{P}$ flips from the $b-$axis into the $a-c$ plane) and x$_{c2}$=0.15 ($\overrightarrow{P}$ flips back from the $a-c$ plane to the $b-$axis). Above x=0.3, the multiferroic state ceases to exist and the only magnetically ordered state is the commensurate and collinear AF4 phase, the same phase as in CoWO$_4$. The change of magnitude and orientation of the ferroelectric polarization as function of x is explained by the rotation of the spin helix driven by the anisotropy of the magnetic ions (Co$^{2+}$ vs. Mn$^{2+}$). The driving mechanism of the multiple polarization flops is suggested to be of thermodynamic origin and it is related to the reduction of the magnetoelectric energy gain, whenever the rotation of the spin helix causes a significant decrease of the polarization. A model relating the magnitude and direction of the polarization to the orientation of the spin helix is derived and provides a qualitative description of the multiferroic properties in the phase diagram. In particular, it is shown that the nearest neighbor intrachain magnetic coupling alone cannot reproduce the experimental results and the longer range interchain interactions have to be taken into account. Between x=0.1 and x=0.15, a new magnetic and paraelectric phase is discovered, sandwiched between the AF4 and AF5 phases. The existence of this phase is the origin of the observed dip in the T$_C$(x) function of the multiferroic phase.

$Note$ $added$: After completion of this work we have been informed about a related work by Urcelay-Olabarria et al. \cite{skumryev:12} on Mn$_{0.8}$Co$_{0.2}$WO$_4$ (x=0.2). The reported results for the magnetization and ferroelectric polarization are in good agreement with the data reported here.

\ack
This work is supported in part by the T.L.L. Temple Foundation, the J.J. and R. Moores Endowment, the State of Texas through TCSUH, the USAF Office of Scientific Research, and at LBNL through the US Department of Energy.

\section*{References}
\bibliographystyle{unsrt}

%\bibliography{HMO}

\end{document}